\documentclass[12pt]{article}
\usepackage{graphicx,times,latexsym,epsfig,rotate,pslatex}

\newcommand{\bit}{\begin{itemize}}
\newcommand{\eit}{\end{itemize}}

\newcommand{\fnaldocheader}[2]{
}

\fnaldocheader{FERMILAB-TM-2279-AD}{November 2004}

\begin{document}

\title{A Sensitivity Study for a MICE Liquid Hydrogen Absorber}

% \title{Radiation Shielding Calculations for MuCool Test Area at Fermilab\thanks{The work was supported by the Illinois
% Board of Higher Education with the Higher Education Cooperative
% Act Grant and Universities Research Association, Inc., under
% contract DE-AC02-76CH03000 with the U. S. Department of Energy. }}

\author{D.~Errede$^{1}$ and I.~Rakhno$^{1,2}$ \\
\normalsize \itshape{$^{1}$Department of Physics, University of Illinois at Urbana-Champaign,}\\
\normalsize \itshape{1110 W. Green St.,  Urbana, IL 61801}\\
\normalsize {\itshape $^{2}$Fermilab, P.O. Box 500, Batavia, IL 60510}
       }

\date{\today}

\maketitle

\begin{abstract}

The International Muon Ionization Cooling Experiment (MICE) is devoted to a study of a muon cooling channel
capable of giving the desired performance for a Neutrino Factory.  One of the goals is achieving
an absolute accuracy of measurements of emittance reduction as high as $\pm$0.1\%.
The paper describes results of a Monte Carlo study on allowed density variations of liquid
hydrogen corresponding to the desired accuracy of the measurements.  

\end{abstract}

\section{Introduction}

An experiment which allows the investigation of the performance of a muon
cooling channel at different conditions of interest has been designed by MICE
collaboration~\cite{MICE_proposal_to_RAL}.  In particular,
high precision measurements of emittance reduction are planned to be performed with an accuracy
of $\pm$0.1\%.  
Various factors should be taken into account to achieve the desired accuracy.  One of the essential
parameters is material density of the liquid hydrogen in the absorber.  The density is influenced 
by temperature and pressure of the hydrogen kept within pre-determined tolerances.  The Monte Carlo
study with the \textsc{MARS14} code \cite{bib:mars} was undertaken to calculate the dependence of the 
emittance reduction on hydrogen density
at \emph{realistic conditions}.  The results can be used to determine the above-mentioned tolerances 
corresponding to the desired accuracy of measurements.

\section{Basic Formulae}

In a four-dimensional phase space with coordinates $x$, $p_x$, $y$, $p_y$, the normalized emittance 
of a muon beam, $\epsilon_n$, can be calculated according to the expression~\cite{Handbook}

\begin{equation}
\label{math/1}
                      \epsilon_n = \pi \sqrt[4]{det\Sigma},
\end{equation}         

\noindent
where $\Sigma$ is a $4\times4$ correlation matrix,
\vspace{1mm}

\begin{equation}
\label{math/2}
             \Sigma = \left( \begin{array}{cccc} \overline{aa} & \overline{ab} & \overline{ac} & \overline{ad} \\
                                                 \overline{ba} & \overline{bb} & \overline{bc} & \overline{bd} \\
                                                 \overline{ca} & \overline{cb} & \overline{cc} & \overline{cd} \\
                                                 \overline{da} & \overline{db} & \overline{dc} & \overline{dd} 
                               \end{array} \right),
\end{equation}  

\vspace{3mm}
\noindent
and $a$ is $x-\bar{x}$, $b$ is $(p_x-\bar{p_x})/m_{\mu}c$, $x$ and $p_x$ are muon coordinate and momentum,
respectively, along $x-$axis.  The entries $c$ and $d$ are for $y-$axis and analogous to $a$ and $b$, respectively.
In the expressions the bar over the symbols means statistical averaging over an ensemble of simulated muon trajectories.

To generate a proper muon beam distribution in the system, symmetry considerations are taken into account~\cite{BVI}.
Firstly, Gaussian distributions are modeled for $x, p_x, y, p_y$ in the geometrical center of the absorber, where
magnetic field flips, using information on the $\beta$-function distribution in the channel~\cite{MICE_proposal_to_RAL}.
Secondly, backward muon transport is modeled in the magnetic field and without the material (hydrogen and aluminum)
to get a reflected incoming muon distribution in front of the absorber.  And thirdly, regular muon transport
through the absorber is modeled. 

For a given ensemble of trajectories, the matrix $\Sigma$ is calculated both in front of and behind the liquid
hydrogen absorber.  After that the emittance reduction, $\Delta\epsilon_n$, is calculated according to expression
(\ref{math/1}).

\section{Geometry Model}

\begin{figure}[htb!]
\vspace{0mm}
\centering\epsfig{figure=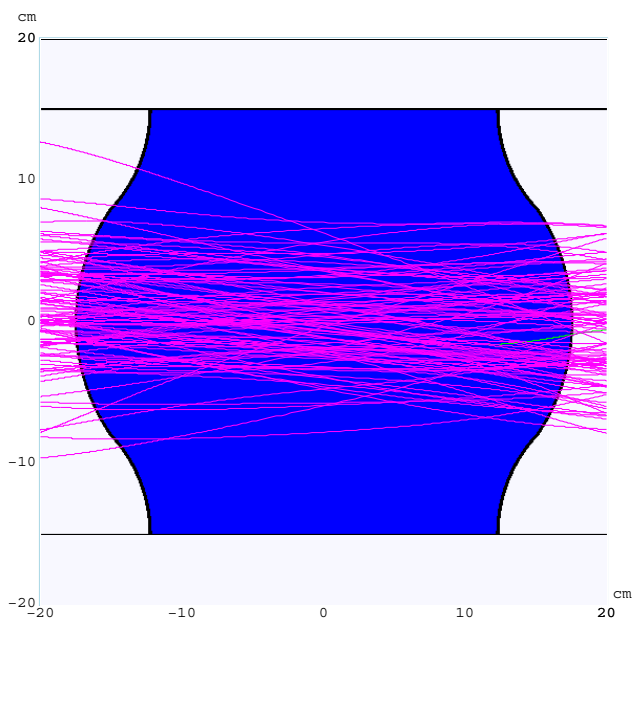,width=0.64\linewidth}
\vspace{-3mm}
\caption
{A cross section of the \textsc{MARS} model of a MICE liquid hydrogen absorber along with 100 sampled muon tracks.}
\label{LH2_absorber_YZ}
\end{figure}

Realistic geometry of the 35-cm liquid hydrogen absorber along with 'inflected' window 
design~\cite{MICE_proposal_to_RAL, Black} was implemented in the model (see Fig.~\ref{LH2_absorber_YZ}).
The absorber is inside a solenoid which is taken into account via its magnetic field only.
Thickness of the aluminum windows of the absorber is variable with the thinnest parts 
being on the $z$-axis.

\begin{figure}[htb!]
\vspace{1mm}
\hspace{-2mm}
\begin{minipage}[t]{0.52\linewidth}
\centering\epsfig{figure=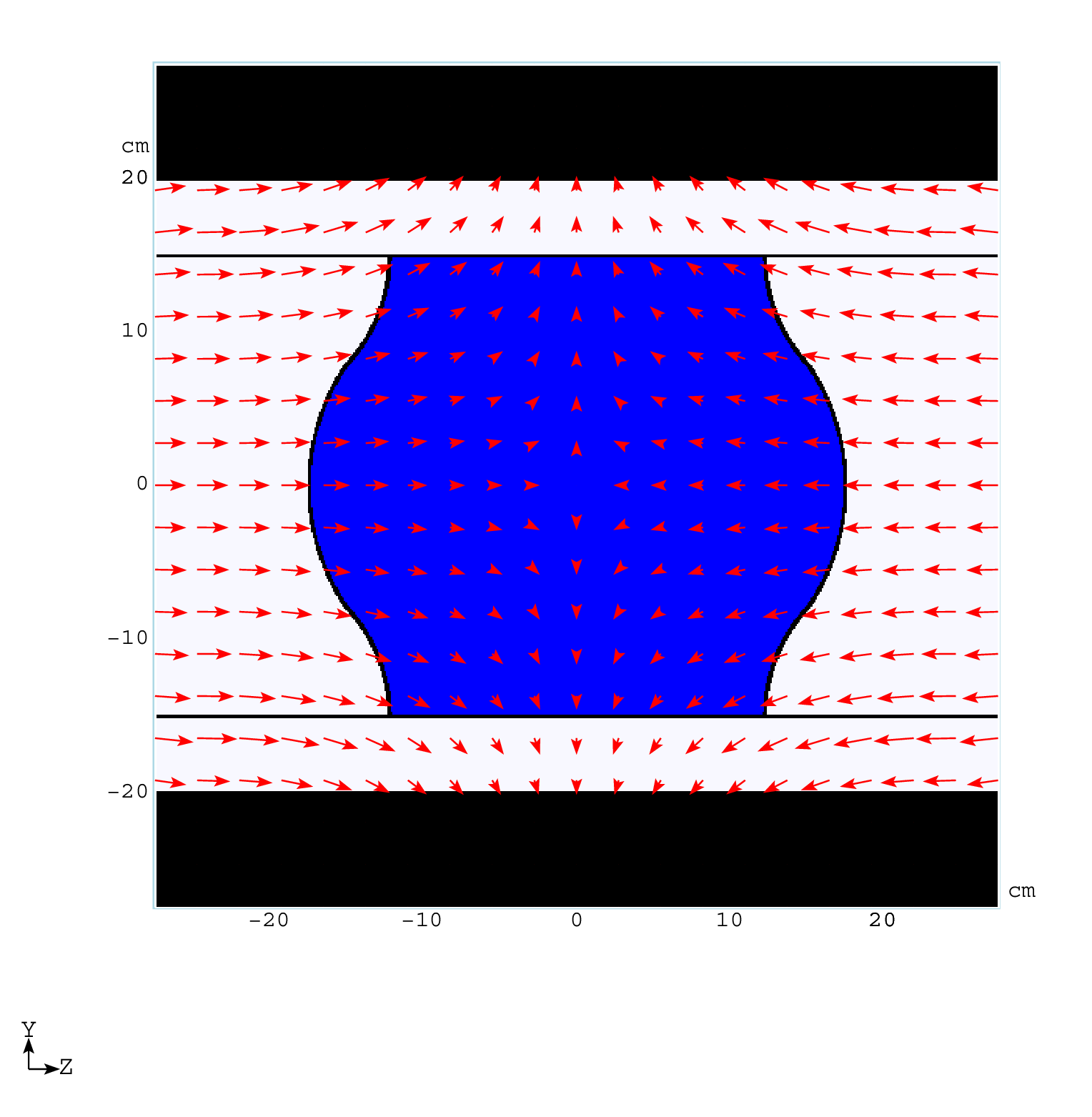,width=\linewidth}
\end{minipage}

\hfill

\vspace{-86.6mm}
\hspace{78mm}
\begin{minipage}[t]{0.52\linewidth}
\centering\epsfig{figure=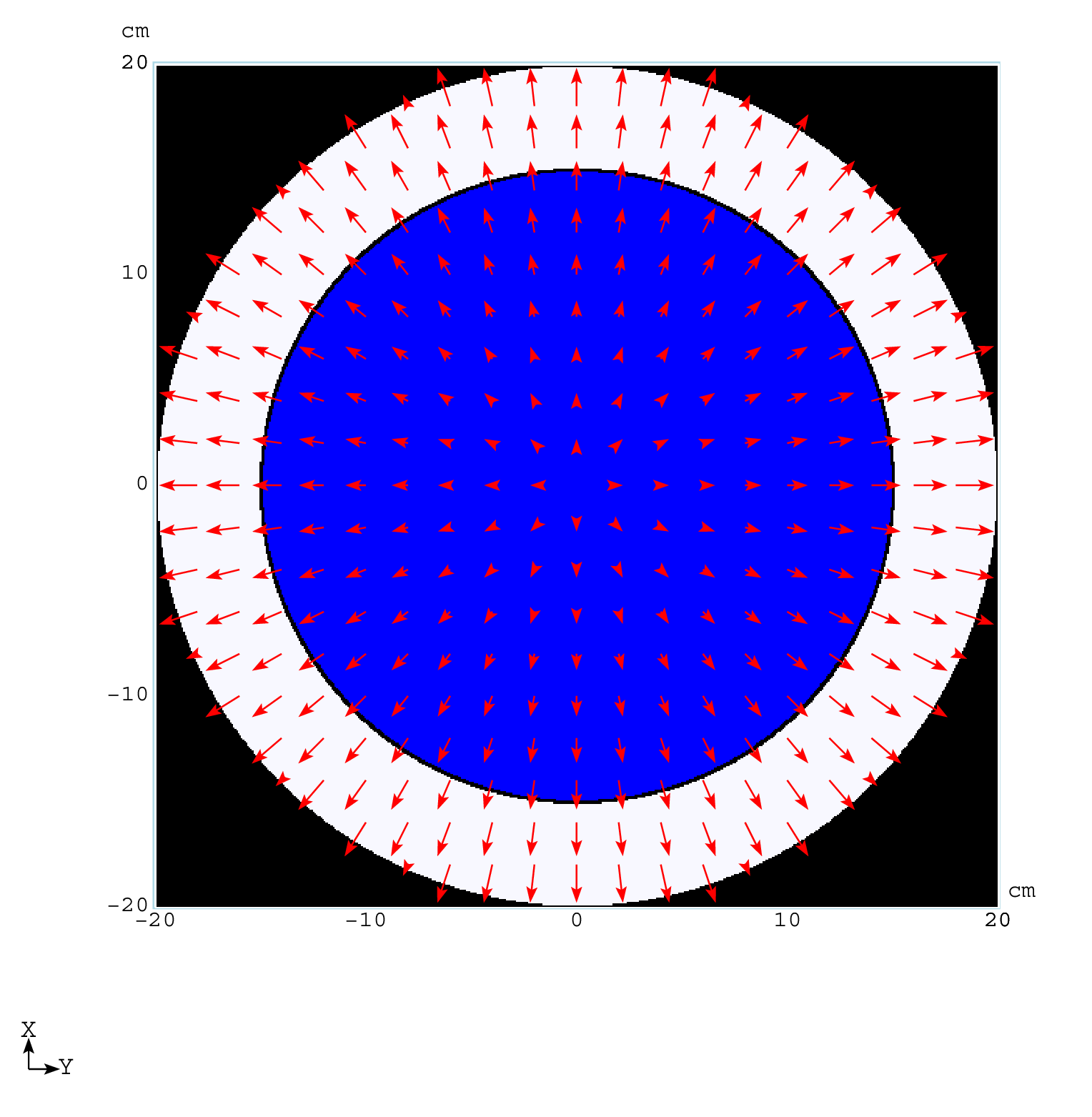,width=\linewidth}
\end{minipage}

\vspace{-4mm}
\caption{The longitudinal (left) and radial (right) distributions of magnetic field around the liquid
hydrogen absorber.  The arrows indicate the field direction only, not magnitude.
The field flips at the geometrical center of the absorber.}
\label{B}
\end{figure}

Realistic three-dimensional distributions of magnetic field over the cooling channel were taken from Ref.~\cite{B_IIT}.
The longitudinal and radial solenoidal field distributions as implemented in the \textsc{MARS} model are shown 
in Fig.~\ref{B}.

\vspace{3mm}

\section{Calculated Results}

\begin{figure}[htb!]
\vspace{-1mm}
\centering\epsfig{figure=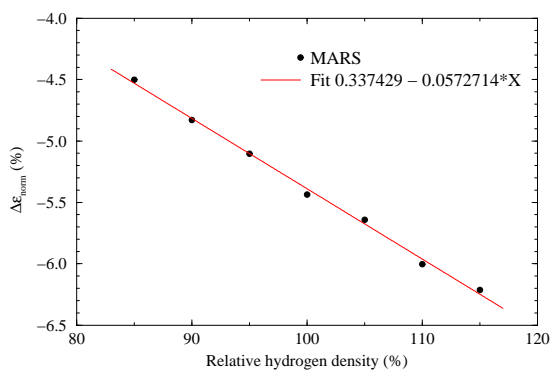,width=0.54\linewidth}
\vspace{-4mm}
\caption
{Calculated emittance reduction \emph{vs} liquid hydrogen density in the central absorber of the 
cooling channel for 200-MeV/c muons~\cite{MICE_proposal_to_RAL}.  Here 100\% corresponds to 0.0708 g/cm$^3$.}
\label{results}
\end{figure}

The Monte Carlo calculations were performed for 200-MeV/c incident muons at various densities of the
liquid hydrogen.  Results of the calculations are shown in Fig.~\ref{results}.
Statistical uncertainty (1$\sigma$) of the calculations was about 0.02\%
which is less than the linear size of the symbols used in the Figure.

One can see that a variation as high as 2\% in hydrogen density gives rise to a 0.1\% variation
in $\Delta\epsilon_n$.  Therefore, to ensure the desired accuracy of emittance measurements, tolerance
for the density of the liquid hydrogen should not exceed 2\%.

\section{Conclusions}

Monte Carlo calculations were performed on emittance reduction of a muon beam \emph{vs} hydrogen density
for a MICE liquid hydrogen absorber within realistic absorber geometry and with detailed three-dimensional
distribution of magnetic field.  It was shown that, within the range of interest, the density dependence
is clearly linear.  To ensure the accuracy of emittance measurements as high as
0.1\%, tolerance for the hydrogen density should be less than 2\%.

\section{Acknowledgements}

The authors are thankful to Valeri Balbekov of Fermilab and Edgar Black of Illinois Institute of Technology
for helpful discussions.

\vspace{3mm}
\noindent
The work was supported by the Illinois
Board of Higher Education with the Higher Education Cooperative
Act Grant and Universities Research Association, Inc., under
contract DE-AC02-76CH03000 with the U. S. Department of Energy.

\end{document}